\documentclass[12pt]{article}
\usepackage{url}
\usepackage{amsmath}
\usepackage{natbib}
\usepackage{amssymb}
\usepackage{graphicx}
\usepackage{epstopdf}

\def\y{\mathbf{y}}

\def\X{\mathbf{X}}

\def\W{\mathbf{W}}
\def\V{\mathbf{V}}
\def\I{\mathbf{I}}

\def\Norm{\text{Normal}}
\def\v{\mathbf{v}}
\def\w{\mathbf{w}}
\def\X{\mathbf{X}}
\def\XX{\mathbb{X}}

\def\E{\mathbf{E}}

\def\b{\mathbf{b}}
\def\B{\mathbf{B}}

\def\z{\mathbf{z}}

\def\Tau{{\boldsymbol{\rm T}}}

\begin{document}
\def\spacingset#1{\renewcommand{\baselinestretch}%
{#1}\small\normalsize} \spacingset{1}

\title{\bf Bayesian predictive modeling of multi-source multi-way data}



  \author{Jonathan Kim$^1$, Brian J. Sandri$^{2,3}$, Raghavendra B. Rao$^{2,3}$, Eric F. Lock$^1$\\ \\
	   $^1$ Division of Biostatistics, School of Public Health \\
	   $^2$ Division of Neonatology, Department of Pediatrics \\
	   $^3$ Masonic Institute for the Developing Brain
	   University of Minnesota}

\maketitle


\begin{abstract}
{ We develop a Bayesian approach to predict a continuous or binary outcome from data that are collected from multiple sources with a multi-way (i.e.. multidimensional tensor) structure.  As a motivating example we consider molecular data from multiple 'omics sources, each measured over multiple developmental time points, as predictors of early-life iron deficiency (ID) in a rhesus monkey model. We use a linear model with a low-rank structure on the coefficients to capture multi-way dependence and model the variance of the coefficients separately across each source to infer their relative contributions.  Conjugate priors facilitate an efficient Gibbs sampling algorithm for posterior inference, assuming a continuous outcome with normal errors or a binary outcome with a probit link. Simulations demonstrate that our model performs as expected in terms of misclassification rates and correlation of estimated coefficients with true coefficients, with large gains in performance by incorporating multi-way structure and modest gains when accounting for differing signal sizes across the different sources. Moreover, it provides robust classification of ID monkeys for our motivating application.  Software in the form of R code is available at \url{https://github.com/BiostatsKim/BayesMSMW}.}
\end{abstract}

\section{Introduction}
\label{sec1}

Technological advancements in biomedical research are producing datasets that are very large and have complex structures. Some data are represented as a multi-way array, also called a tensor, which extends the two-way data matrix to higher dimensions. Some data are multi-source, which involves features from different sources of data matched by samples (this is also known as multi-view data). A growing number of datasets are simultaneously multi-source and multi-way (MSMW). 
As a motivating example of MSMW data, we consider predictors of early-life iron deficiency (ID) in infant monkeys using data described in \cite{sandri2022multiomic}. In this naturalistic ID model in infant rhesus monkeys, $20-30\% $ of infants develop ID and anemia between 4 and 6 months of age due to a combination of lower iron stores at birth and rapid postnatal growth rate \citep{lubach2006preconception,coe2013optimal}. Prior studies in this model have shown that the ID infants have metabolomic and proteomic abnormalities in the serum and cerebrospinal fluid in the preanemic and anemic periods with residual changes persisting even after the resolution of anemia with iron treatment \citep{geguchadze2008csf,coe2009history,patton2012quantitative,rao2013metabolomic,rao2018metabolomic,sandri2020early,sandri2021correcting,sandri2022multiomic}.
Data were available from two sources, serum proteomics and serum metabolomics, collected at two time points, 4 and 6 months after birth. The data therefore form two 3-way arrays: [monkeys $\times$ proteomics  $\times$ time] and [monkeys $\times$ metabolomics  $\times$ time]. This motivating data is therefore MSMW and we are interested in identifying signals in the biomarkers that can predict ID status. 

To understand the significance of incorporating MSMW structure into analysis, consider a naive approach in which each source's multi-dimensional data array is transformed into a vector and features from all sources are concatenated into a single vector. While this approach would produce data that could be analyzed using one of the many methods available for vector-valued data, it would also have a number of shortcomings. Ignoring the multi-way structure would not allow for consideration of dependence across dimensions. Ignoring the multi-source structure would mean that any signal present in features from smaller sources could be overrun by noise from larger sources with comparatively less signal.

A common aspect of MSMW data is the presence of far more features than samples, often referred to as high-dimension low-sample size (HDLSS) data. While MSMW data need not necessarily be HDLSS, it is sufficiently common that methods for handling MSMW data will ideally allow for HDLSS structure. A Bayesian framework provides more flexibility for model-based supervised analysis of high-dimensional data, as appropriate regularization can be induced through the specified prior distribution.

In what follows we briefly review existing methods for predictive modeling of  data that are multi-source (Section \ref{sec:MSIntro}) and multi-way (Section \ref{sec:MWIntro}); our methodological contributions are summarized in Section \ref{sec:MSMWIntro}. 

\subsection{Multi-source data}
\label{sec:MSIntro}

The issue of integrating data from multiple sources has been addressed in a variety of ways for different tasks. 
For predicting an outcome from multi-source data, several approaches extend unsupervised methods that were originally designed to integrate multi-source data without prediction. Examples include various supervised extensions of canonical correlation analysis (CCA). \citet{rodosthenous2020integrating} extend CCA to an arbitrary number of sources and an outcome by means of a generalized sparsity parameter.  
Joint association and classification (JACA) \citep{zhang2021joint} is a combination of CCA and linear discriminant analysis for a binary outcome. Data integration analysis for biomarker discovery using latent components (DIABLO) is an extension of both sparse projection to latent structure discriminant analysis to multi-omics analyses and sparse generalized CCA to a supervised analysis framework by \cite{singh2019diablo}, and a combination of multivariate ANOVA with Bayesian CCA was developed by \cite{huopaniemi2010multivariate}. Supervised JIVE (sJIVE) \citep{palzer2021sjive} was developed as a supervised extension for prediction using the JIVE method, which decomposes data into latent factors that are shared or specific to each source.

Other methods can be used directly in a supervised context (i.e., classification and regression), without incorporating aspects of unsupervised analysis.
One approach developed by \cite{van2016better}  handles 
``co-data", which they defined as ``all information on the measured variables other than their numerical values for the given study". In particular, their method involved partitioning variables into groups and imposing group-specific penalties for ridge regression. This approach has some analogues to the multi-source problem in that it is able to perform prediction on a binary or continuous outcome using data from multiple groups; though such ``groups" of data included p-values from previous studies or genomic annotations, 
in the multi-source context they may be defined by which source the variable belongs to (e.g., proteins or metabolites).
A Bayesian approach to multi-source data that make use of the prior distribution to accommodate different sources has been used recently by \cite{white2021bayesian}. However, their Bayesian Multi-Source Regression (BMSR) assumed double-matched multi-source data, i.e., the same features present across all sources, and involved predicting different outcomes for each source instead of a single outcome affected by multiple sources.
 
One limitation of all of these multi-source methods is that they do not have the ability to accommodate data that exists in multiple ways, thereby limiting their physiological importance to that trough of data potentially limiting critical findings.

\subsection{Multi-way data}
\label{sec:MWIntro}

Methods developed for analyzing data with multi-way structure can be divided into unsupervised and supervised categories, with the latter further divided between classification and regression methods. Unsupervised approaches to handling multi-way data predominantly involve dimension reduction techniques, which reduce the number of features in the data to a more manageable size while preserving the overall integrity of the data. Many such methods of tensor decomposition are outlined in \cite{kolda2009tensor}.
\cite{gloaguen2022multiway} developed a multi-way extension of regularized generalized canonical correlation analysis that can accommodate data with a tensor structure from an arbitrary number sources by incorporating Kronecker constraints into the optimization problem.

For supervised methods involving classification, there is growing literature on extending classifiers of vectors to multi-way arrays using factorization and dimension reduction techniques. \cite{tao2005supervised} proposed a supervised tensor learning framework that generalized classifiers by performing a rank-1 decomposition on the coefficients to reduce their dimension to a single set of weights for each dimension. \cite{lyu2017discriminating} proposed a multi-way version of the classification method distance weighted discrimination (DWD) under the assumption that the coefficient array is low-rank. Their implementation of multi-way DWD was shown to dramatically improve performance over two-way classifiers when the data have multi-way structure. However, their method is restricted to use for three-way data. \cite{guo2021multiway} proposed an extension of multi-way DWD that also imposed a low-rank structure on the coefficient array, but allowed for data with an arbitrary number of ways and accounted for sparsity. 

Supervised methods of regression also build on dimension reduction techniques by extending them to the regression context. Both \cite{zhou2013tensor} and \cite{li2018tucker} propose maximum likelihood estimation algorithms that could perform regression with array-valued covariates through dimension reduction, with the former using CANDECOMP/PARAFAC (CP) decomposition and the latter using Tucker decomposition. A Bayesian formulation of tensor regression was developed by \cite{miranda2018tprm}, which involves a multi-step process of partitioning tensor data into smaller sub-tensors, reducing these sub-tensors via CP decomposition, and performing regression with sparsity-inducing priors to identify informative sub-tensors. Another Bayesian approach to tensor regression with a scalar response was developed by \cite{guhaniyogi2017bayesian} by means of a novel multi-way shrinkage prior, which allows for simultaneous shrinkage of parameters across all ways of data.

In the same way that existing multi-source methods have yet to be extended to accommodate multi-way data, existing supervised multi-way methods are generally unable to incorporate data from multiple sources.  

\subsection{Contributions: method for multi-source multi-way data}
\label{sec:MSMWIntro}

In this paper, we develop a Bayesian linear model that can perform regression or classification on MSMW data for either a continuous outcome with normal errors or a binary outcome with a probit link, respectively. The central assumption for our multi-way approach is that the signal discriminating the ways can be efficiently represented by meaningful patterns in each dimension, which we identify by imposing a low-rank structure on the coefficient array. The central assumption for our multi-source approach is that the signal discriminating the sources can be efficiently represented by 
modeling the variances of the coefficients separately across each source to infer their relative contributions. We incorporate both of these approaches into a single model under a Bayesian framework. We also apply our method to a real-world MSMW dataset by predicting iron deficiency status in infant monkeys based on multi-omic tissue samples. 


\section{Methods}
\label{sec2}

\subsection{Notation and framework}
\label{sec:notation}

Throughout this article bold lowercase characters ($\mathbf{a}$) denote vectors, bold uppercase characters ($\mathbf{A}$) denote matrices, and blackboard bold uppercase characters ($\mathbb{A}$) denote multi-way arrays of the specified dimension (e.g., $\mathbb{A} : P_1 \times P_2 \times \cdots \times P_K $). Square brackets index entries within an array, e.g., $\mathbb{A} \big[ p_1, p_2, \ldots, p_K \big]$. Superscripts in square brackets are used to denote individual sources within a multi-source set of data, e.g., $\mathbb{A}^{[1]},\mathbb{A}^{[2]}, \ldots, \mathbb{A}^{[M]}$.  Define the generalized inner product for two arrays $\mathbb{A}$ and $\mathbb{B}$  of the same dimension is 
\[\mathbb{A} \cdot \mathbb{B} = \sum_{p_1=1}^{P_1} \cdots \sum_{p_K=1}^{P_K} \mathbb{A}[p_1,\hdots,p_K] \mathbb{B}[p_1,\hdots,p_K].\] 
Define $||\cdot||_F$ as the Frobenius norm and $\text{vec}(\mathbb{A})$ as the vectorization of the entries in $\mathbb{A}$.  

For our context, $\XX: N \times [P^{[1]},\cdots,P^{[M]}] \times D$ gives data in the form of a 3-way array for $N$ subjects, where $P^{[m]}$ is the size of the $m^{th}$ source for $m = 1, . . . , M$ with $P=\sum_{m=1}^M P^{[m]}$, and $D$ is an additional way for which we have data available for all subjects and sources. Each subject has a response variable $y_i$, which may be binary or continuous; let $\y = [y_1, . . . , y_N]$. Our goal is to predict the outcome $\y$ based on the multiway covariates $\XX$.


\subsection{Model}

\subsubsection{Bayesian linear model}
\label{sec:LM}


We first briefly consider the special case in which we have only one way of data from only one source, $M=1$ and $D=1$, and $\y$ is continuous.  This is the classical setting with $P$ covariates available for $N$ subjects, $\X: N \times P$. 
The basic linear model is $\y = \X^T \mathbf{b} + \mathbf{e}$ where $\mathbf{b} = 
[b_1,...,b_P]$
is the vector of covariate coefficients and $\mathbf{e} = [\epsilon_1,...,\epsilon_N]$ is the vector of error terms, which are assumed to have distribution $\epsilon_i \stackrel{iid}{\sim} \Norm(0,\sigma^2)$. 

Under a Bayesian framework, we place a prior distribution on $\b$ and we assume all $b_j, j=1,...,P$ are independent and identically distributed under this prior. If we let that prior be a normal distribution with mean zero and 
variance $\tau$, i.e. $b_j \sim \Norm(0,\tau)$, we can also place a hyperprior on the variance $\tau$.  
This has the advantage of empirically controlling the level of shrinkage of the coefficient toward $0$, via the posterior for $\tau$.   In subsequent sections we extend this model to the multi-source and multi-way scenarios.  



\subsubsection{Multi-source model}
\label{sec:MSLM}

Now, suppose our covariates come from $M>1$ different sources, which can be conceived as $M$ datasets $\X_1,...,\X_M$ with each $\X_m$ being an $N \times P^{[m]}$ matrix such that $\sum_{m=1}^{M}P^{[m]} = P$. The vector of coefficients, $\b$, can be represented as a concatenation of $M$ vectors each of length $P^{[m]}$, i.e. $\mathbf{b} = [\b^{[1]},...,\b^{[M]}]$ with each $\b^{[m]} = [b_1^{[m]},...,b_{P^{[m]}}^{[m]}]$ for $m = 1,...,M$.

To infer the relative contribution of each source, we propose modeling the variances of each source's coefficients separately such that each source has its own independent prior placed on its coefficients. That is, let $\b^{[m]} \sim D_m$ where $D_m$ is an arbitrary distribution. If we let the priors all be mean-zero normal distributions as in the previous section, that gives us
\begin{align}
\b^{[m]} \sim MVN(\mathbf{0},\tau_m \I). \label{eq:MSmodel}
\end{align}
This allows us to distinguish the 
level of the contribution for coefficients from different sources by allowing for different source variances $\tau_m$.


\subsubsection{Multi-way model}
\label{sec:MWLM}

\subsubsection*{Rank 1 model}

We now consider the case in which the data are multi-way ($D>1$) but single source $P=1$, and thus $\XX$ is $N \times P \times D$, a 3-way tensor. We propose the following bilinear model, which is analogous to the one proposed by \cite{lyu2017discriminating} in the context of modeling the coefficients for multi-way DWD.   
 
We assume the covariate matrix $\mathbf{B}: P\times D$ has rank-1 decomposition 
\begin{align}
\mathbf{B}=\w \v^T \label{eq:rank1B}
\end{align}
where $\w=[w_1,...,w_P]^T $ and $\v=[v_1,...,v_D]^T$.

Thus our model for each $i \in 1,...,N$ is
\begin{align}
y_i = \X_i \cdot \mathbf{B} = \v \X^T_i\w \label{eq:Model}
\end{align}
where $\v$ is $1 \times D$, $\X_i$ is $P \times D$, and $\w$ is $P \times 1$.

To interpret this model in the context of our motivating example, we may consider $\w$ to represent a pattern in the metabolites that is predictive of $y$ while $v$ gives the relative contribution at each time point. 

\subsubsection*{Rank R model}

In the previous model \eqref{eq:rank1B}, we assumed the covariate matrix $\mathbf{B}$ had a rank-1 decomposition, that is, the outcome is determined by combining a single pattern in each dimension of the coefficient matrix. However, it is possible that multiple patterns contribute to the outcome. For example, it may be that some metabolites are predictive of the outcome at an early time point but others are only predictive of the outcome at a later time point. Consider a new data structure where we assume the covariate matrix $\mathbf{B}$ has rank-R decomposition: 
\begin{align}
\B = \W \V^T \label{eq:rankRB}
\end{align}
where $\W: P \times R$ with columns $\w_r=[w_{r1},...,w_{rp}]^T $ and $\V: D \times R$ with columns $\v_r=[v_{r1},...,v_{rd}]^T$, for $r = 1,...,R$, $R < min(P,D)$. Observe that the coefficient matrix $\mathbf{B}$ in the rank-1 multi-way model \eqref{eq:rank1B} is a special case of the rank-R multi-way model \eqref{eq:rankRB} when $R = 1$. 

This use of low-rank structure on the coefficients allows us to capture multi-way dependence and identify relevant patterns in each dimension of the coefficient matrix. 

\subsubsection{Multisource and multi-way model}
\label{sec:MSMWLM}

We now combine aspects of the multi-source and multi-way models into a single Bayesian linear model 
to address the general MSMW framework introduced in Section~\ref{sec:notation}.
We again assume the covariate matrix $\mathbf{B}$ has rank-R decomposition as in \eqref{eq:rankRB}
%
where $\W= [\W^{[1]},...,\W^{[M]}]$ and each $\W^{[m]}$ has columns $\w_r^{[m]} = [w_{r1},...,w_{rp}]^T $ for $m = 1,...,M$ and $\V$ has columns $\v_r=[v_{r1},...,v_{rd}]^T$for $r = 1,...,R$, $R < min(P,D)$.

\subsubsection{Binary outcome}
\label{sec:Binary}

We now consider our model in the case where $\textbf{y}$ is binary, i.e. $y_i \in \{0,1\}$ for $i = 1,...,N$.  We can accommodate such data by modifying our approach to use a latent variable probit model, similar to that described in \cite{albert1993bayesian}. Suppose there exists an auxiliary random variable $z_i$ such that $z_i = \X_i \cdot \B + \epsilon_i$, where $\epsilon_i \sim \Norm(0,1)$. We can model our outcome variable $y_i$ as an indicator for whether or not this latent variable is positive, that is, $y_i = 1$ if $z_i >0$ and $y_i = 0$ otherwise. This is equivalent to using a probit link function:
$$Pr(y_i=1|\X_i) = Pr(z_i > 0 | \X_i) = Pr(\X_i \cdot \B + \epsilon_i >0) = Pr(\epsilon_i < \X_i \cdot \B) = \Phi(\X_i \cdot \B)$$
where $\Phi(x) = \frac{1}{\sqrt{2\pi}} \int_{-\infty}^{x}e^{\frac{-t^2}{2}}dt $.

\subsection{Model estimation}

\subsubsection{Priors}
\label{sec:priors}

As referenced in Section \ref{sec:LM}, we can model the coefficients in a single-source non-multi-way model to come from a normal distribution with mean zero and variance $\tau$. We extend this approach to our MSMW model by using a mean-zero normal prior to estimate the components in our covariate matrix $\B$.

In order to accommodate our multi-way model, we do not estimate $\B$ directly, but instead estimate the components of our covariate matrix, either $\w$ and $\v$, as outlined in \eqref{eq:rank1B} for the rank-1 model or $\W$ and $\V$ as outlined in \eqref{eq:rankRB} for the rank-R model. We then place mean-zero normal priors on each $\W$ and $\V$, that is, 
the entries of each $\W^{[m]}$ are independent with a $\Norm(0, \tau_m)$ distribution, and the entries of $\V$ are independent with a $\Norm(0,1)$ distribution.   We fix the variance of $\V$ because $\B$ is the product of $\V$ and $\W$, and thus their respective scales are not identifiable and only the variance of $\W$ needs to be modeled.  This further allows us to model the variance of the contribution for each source separately by considering each $\tau_m | \W^{[m]}$ for $m = 1,...,M$.  In order to facilitate an efficient sampling algorithm for the posterior distribution, we place conjugate inverse-gamma priors on the variance parameters, $\tau_m \sim IG(\alpha_0,\beta_0)$.   

We assume that the error terms $\mathbf{e}$ are independent and normally distributed, $\mathbf{e} \sim MVN(0,\I \sigma^2)$. For a continuous $\y$, $\sigma^2$ may either be fixed or given a prior. If $\sigma^2$ is unknown, by default we use an inverse-gamma prior distribution with arbitrarily small hyperparameters as a non-informative prior, e.g., $\sigma^2 \sim IG(0.001,0.001)$.  If $\y$ is binary, then the error variance for the latent continuous variables $\z$ in Section~\ref{sec:Binary} is fixed at $\sigma^2=1$.

\subsubsection{Full conditional distributions}
\label{sec:conditionals}

Given the conjugate hyperpriors we've placed on $\tau$, and fixing the variance of our error terms at 1, we will have the following conditional distribution for each $\tau_m$:
\begin{align}
\tau_m | \W \sim IG \left(\alpha_0+\frac{P^{[m]}}{2}, \beta_0 + \frac{1}{2} ||\W^{[m]} ||_F^2 \right). \label{eq:tauwconditional}
\end{align}



For our coefficient factor parameters, $\W$ and $\V$, standard linear model results with conjugate normal priors \citep{lindley1972bayes} produce:  
\begin{align}
\text{vec}(\W) |\y,\boldsymbol{\tau},\V, \sigma^2 \sim MVN((\Tau^{-1}\sigma^2 + \X_v^T\X_v)^{-1}( \X_v^T \y), \sigma^2(\Tau^{-1}\sigma^2+\X_v^T\X_v)^{-1}) \label{eq:MWwconditional}
\end{align}
where $\Tau: RP \times RP$ is the diagonal prior covariance matrix with diagonal entries
\[[\underbrace{\tau_1 \, \hdots \tau_1}_{P^{[1]}} \,  \underbrace{\tau_2 \hdots \tau_2}_{P^{[2]}} \, \hdots \, \underbrace{\tau_M \hdots \tau_M}_{{P^{[M]}}}]\] repeated $R$ times, and $\X_v: N \times RP$ is the matrix with row $i$ given by $\text{vec}(\X_i \V)$.  Similarly, 
\begin{align}
\text{vec}(\V) |\y,\W, \sigma^2 \sim MVN((\I\sigma^2 + \X_w^T\X_w)^{-1}( \X_w^Ty), \sigma^2(\I\sigma^2+\X_v^T\X_v)^{-1}) \label{eq:MWvconditional}
\end{align}
where $\X_w: N \times RD$ is the matrix with row $i$ given by $\text{vec}(\W \X_i)$.

Use of the non-informative conjugate prior ($\sigma^2 \sim IG(0.001,0.001)$) yields the  following full conditional distribution for $\sigma^2$:
\begin{align}
\sigma^2 | \y, \B \sim IG \left(\frac{N}{2} + 0.001, \frac{(\y-\XX\B)^T(\y-\XX\B)}{2} + 0.001 \right). \label{eq:sigmaconditional}
\end{align}

\subsubsection{Data augmentation of binary case}
\label{sec:augment}

Under the latent variable formulation for binary data described in Section~\ref{sec:Binary}, the full conditional distributions for $\W$ and $\V$ are analogous to that in are analogous to that in \eqref{eq:MWwconditional} and \eqref{eq:MWvconditional}, respectively, but with $\z$ replacing $\y$ and $\sigma^2=1$.  



As a consequence of our latent variable modeling, the conditional distribution of $\z$ will be a truncated normal distribution, denoted as $N_{trunc}$, as follows: 
\begin{align}
z_i |\y,\B \sim N_{trunc}(\X_i \cdot \B,1) \label{eq:truncnorm}
\end{align}
where the distribution is truncated at the right by 0 if $y_i = 0$ and truncated at the left by 0 if $y_i = 1$. 

\subsubsection{Gibbs sampling algorithm for continuous case}
\label{sec:continuousgibbs}

We approximate our posterior using a Gibbs sampling algorithm. Here we provide the algorithm used for the continuous version of the MSMW model to draw samples from the joint posterior distribution $p(\W,\V, \boldsymbol{\tau}, \sigma^2|\XX,\y)$. The algorithm is given below for iterations $t=1,...,T$: 

      

\begin{enumerate}
  \item Initialize $\W^{(1)},\sigma^{(1)}, \tau_1^{(1)},...,\tau_M^{(1)}$
  \item Make the following draws for $2,...,T$
  \begin{itemize}
      \item Draw $\V^{(t)} | \y,\sigma^{(t-1)}, \W^{(t-1)}$ as in \eqref{eq:MWvconditional}
      \item Draw $\W^{(t)} | \y,\sigma^{(t-1)},\tau_1^{(t-1)},\tau_2^{(t-1)}, \V^{(t)}$ as in \eqref{eq:MWwconditional}
      \item Draw $\tau^{(t)}_1,...,\tau^{(t)}_M | \W^{[1](t)},...,\W^{[M](t)}$ as in \eqref{eq:tauwconditional}
      \item Calculate $\B^{(t)} = \W^{(t)} \V^{(t)^T}$
      \item Draw $\sigma^{(t)}| \B, \y$ as in \eqref{eq:sigmaconditional} (if $\sigma^2$ is not fixed). 
      
  \end{itemize}
\end{enumerate}

\subsubsection{Gibbs sampling algorithm for binary case}
\label{sec:binarygibbs}

In order to accommodate binary data in our Gibbs sampler, we must introduce our data augmentation steps, in which we draw the latent continuous variables $z_i$:

\begin{enumerate}
  \item Initialize $\V^{(1)},\W^{(1)},\z^{(1)}, \tau_1^{(1)},...,\tau_M^{(1)}$
  \item Make the following draws for $2,...,T$
  \begin{itemize}
      \item Draw $\V^{(t)} | \z, \W^{(t-1)}$ as in \eqref{eq:MWvconditional}, with $\z$ replacing $\y$ and $\sigma^2=1$.
      \item Draw $\W^{(t)} | \y, \z^{(t)},\boldsymbol{\tau}^{(t-1)}, \V^{(t)}$ as in \eqref{eq:MWwconditional}, with $\z$ replacing $\y$ and $\sigma^2=1$. 
      \item Draw $\tau^{(t)}_1,...,\tau^{(t)}_M | \W^{[1](t)},...,\W^{[M](t)}$ as in \eqref{eq:tauwconditional}
      \item Calculate $\B^{(t)} = \W^{(t)} \V^{(t)^T}$.
      \item Draw $\z | \y, \B^{(t)}$ as in \eqref{eq:truncnorm}.
      
  \end{itemize}
\end{enumerate}

\subsubsection{Model Prediction}
\label{sec:prediction}

After running our Gibbs sampler to simulate draws from our posterior, we take the average over sampling iterations $\B^{(1)},...,\B^{(T)}$ to obtain estimated coefficients $\hat{\B}$.  Given new data $\XX^*$ for $N^*$ samples, we can then obtain a point estimate the outcomes $\y^*$ via $\hat{y}_l^*=\X_l^* \cdot \hat{\B}$.  For a binary outcome, $\Phi(\X^*_l \cdot \hat{\B})$ gives an estimate of the predicted probability of having an outcome value of 1, and to translate this probability into a class prediction, we can simply round the value to the nearest integer:

\begin{equation}
\hat{y}_l^*=
    \begin{cases}
        1 & \text{if } \Phi(\X^*_l \cdot \hat{\B}) \geq 0.5\\
        0 & \text{if } \Phi(\X^*_l \cdot \hat{\B}) < 0.5
    \end{cases}
    \label{eq:Prediction}
\end{equation}
for $l = 1,...,N^*$.



Alternatively, the Bayesian approach allows one to model the full posterior predictive distribution with uncertainty.  For the continuous case,  draws $y_l^{*(t)}$ from the posterior predictive distribution can be obtained from the Gibbs draws via  $y_l^{*(t)} \sim \text{Normal}(\X^*_l \cdot \B^{(t)}, \sigma^{2 (t)})$ for $l = 1,...,N^*$.  In the binary case,  draws from the posterior predictive can be generated via  $y_l^{*(t)} \sim \text{Bernoulli}(\Phi(\X^*_l \cdot \B^{(t)}))$ for $l = 1,...,N^*$.

\section{Results}
\label{sec3}

\subsection{Simulations}



\subsubsection{Data generation}
\label{sec:simgenerate}

We generated data under multiple scenarios to illustrate the relative benefits of incorporating multi-source or multi-way structure under different conditions. For all  scenarios, we simulated data sets $\XX^{[1]}: N \times P_1 \times D$ and $\XX^{[2]}: N \times P_2 \times D$, representing data from two sources with $N$ observations, $P_1$ and $P_2$ covariates from each source with $P_1=P_2=P/2$, and $D$ time points. 
We consider a low-dimensional scenario with $N=100$ and $P=6$ and $D=5$, and a high-dimensional scenario with $N=20$ and $P=200$ and $D=2$ (closely matching the application in Section~\ref{sec:Application}).  
We generated the true coefficient array $\B$ under one setting for which the sources contribute equally ($\tau_1=\tau_2=1$) and on setting for which only one source contributes $(\tau_1=0, \tau_2=1)$.   
We also consider settings under which $\B$ has a rank $1$ or rank $2$ decomposition,  or where the coefficient matrix has no multiway structure (i.e., $\B$ has independent entries and is of full rank).   In the non-multiway case, 
the entries of the coefficients for each source $\B^{[m]}:P^{[m]} \times D$ are generated independently from a $\Norm(0, \tau_m)$ distribution. In the multi-way case for rank $R=1$ or $R=2$, 
we generate $\W^{[m]}: P^{[m]} \times R$ and $\V: D \times R$ by simulating 
the entries of each $\W^{[m]}_j$ independently from $\Norm(0, \tau_m)$ for $m=1,2$ and the entries of $\V$ independently from $\Norm(0, 1)$.  A rank 2 model was not considered for the high-dimensional case, because the full rank scenario is already of rank 2 ($D=2$).

\subsubsection*{Continuous outcome}
For the continuous case, the entries of $\XX^{[1]}$ and $\XX^{[2]}$ were each generated independently from a $\Norm(0,1)$ distribution.  Then, after generating $\B$, the response variables $\y$ was generated via $y_i \sim \Norm(\X_i \cdot \B, 1)$. 

\subsubsection*{Binary outcome}
For our first binary data generating procedure, similarly to the continuous case, the entries of $\XX^{[1]}$ and $\XX^{[2]}$ were each generated independently from a $\Norm(0,1)$ distribution.  
Then, after generating $\B$, the response variables $\y$ was generated via using the probit link function $y_i \sim \text{Bernoulli}(\Phi(\X_i \cdot \B))$.

\subsubsection*{Separate normal distributions}
We considered a third case for which the outcome is binary and the distribution of $\X_i$ depends on the outcome. Here, the 
outcome was generated deterministically, with half of the $N$ observations having value $0$ and half having value $1$: $y_i=0$ for $i=1,\hdots,N/2$ and $y_i=1$ for $i=N/2+1,\hdots,N$.  The coefficients $\B$ are generated under the same conditions above, and then $\XX$ is generated via $\X_i=-\B+\E_i$ if $y_i=0$ and $\X_i=\B+\E_i$ if $y_i=1$, with the entries of $\E_i$ generated independently from  a $\Norm(0,1)$ distribution for $i=1,\hdots,N$.
%
Note that this scenario does not explicitly match the assumptions of our probit model, however, it approximates a realistic scenario for which the data have different means depending on their class, which is detectable in the high dimensional case. The estimated coefficients for the optimal linear classifier will be proportional to $\B$.  

\subsubsection{Measures of performance considered}
\label{sec:perform}

We assess predictive performance by applying our model to test data that were generated from the same distributions as the training data with a larger sample size ($N^*=500$).  For our simulations with a binary outcome, we used the prediction method outlined in \eqref{eq:Prediction} and compare the predicted classification to the true classification to get a misclassification rate.  For our simulations with a continuous outcome, we compare the predicted outcome with the true outcome by calculating the relative mean squared error $||\y-\hat{\y}||_F^2/||\y||_F^2$.
For all simulations, we also assess recovery of the underlying parameters by considering the posterior coverage rates of the true parameters and also the correlation between the estimated estimated and the true coefficients. 




\subsubsection{Models used for estimation}

For each simulation condition, we ran a total of six different models that each made different assumptions about the underlying structure of $\B$.  

For non-multiway models, the data were assumed not to follow a multi-way structure and the data arrays were reorganized into a matrix of dimension $N \times PD$ where the $ith$ row gives $\text{vec}(\X_i)$. We also ran two models that did assume a multi-way structure; the first of these models imposed the assumption of a rank 1 covariate coefficient matrix structure (as in Equation \ref{eq:rank1B}) and the second of these models imposed the assumption of a rank 2 covariate coefficient matrix structure (as in Equation \ref{eq:rankRB} for $R = 2$). 

For the multi-source models, the covariates were assumed to come from two sources, with half of the covariates from one source and half from the other. This means that two independent priors on the covariate coefficients were fit as in \eqref{eq:MSmodel} for $m = 1,2$. For the single-source models, the data were assumed to come from one source (i.e., distinction between $\XX_1$ and $\XX_2$ were ignored) and only one prior was placed on the covariate coefficients. 

Taking all combinations of these models produces the following six that were fit in our simulations:

\begin{enumerate}
\item Rank 2, Multi-source model (Rank2,MS) with $\tau_m \sim IG(1,\sqrt{P^{[m]}*R})$ for $m=1,2$
\item Rank 2, Single-source model (Rank2,SS) with $\tau \sim IG(1,\sqrt{P*R})$
\item Rank 1, Multi-source model (Rank1,MS), with  $\tau_m \sim IG(1,\sqrt{P^{[m]}*R})$ for $m=1,2$
\item Rank 1, Single-source model (Rank1,SS) with  $\tau \sim IG(1,\sqrt{P})$
\item Non-multi-way, Multi-source model (FullRank,MS) with  $\tau IG(1,\sqrt{P^{[m]}*d})$ for $m=1,2$ 
\item Non-multi-way, Single-source model (FullRank,SS), with  $\tau \sim IG(1,\sqrt{P*d})$.
\end{enumerate}

\subsection{Simulation results}
\label{sec:results}

The following tables show the results of the metrics outlined in \ref{sec:perform} for all of our simulations, averagedd over $100$ replications for each condition. For these tables, ``MS" is an abbreviation for ``multi-source" and is used to indicate either a multi-source model or a simulation condition in which the signal was equal across both sources (as opposed to being entirely confined to a single source). ``Rank" in these tables refers to the rank of the true coefficient matrix, with ``FullRank" referring to the simulation condition in which the true coefficient matrix is generated without any multiway structure.
For all tables, bolded values indicate the best performing model based on a pairwise t-test. If multiple values are bolded, then model performances were not significantly different at a 0.05 level.

\subsubsection*{Binary outcome, low dimensions (N=100, P1=3, P2=3, d=5)}

\subsubsection*{Probit generated data}
Tables \ref{tab:probit_misclass_ld} and \ref{tab:probit_correlation_ld} give the misclassification rate and correlation with the true discriminating signal, respectively, for the low dimensional simulation with probit generated data.  In all cases, the model that best matched the data generation scenario performed the best for both measures.  The benefits of using the correct multiway structure (Rank 2, 1 or full rank) tended to be more dramatic than that for matching the multi-source structure; the relative differences are particularly large for the correlations with the true coefficients shown in Table~\ref{tab:probit_correlation_ld}. There were a few cases in which a model did not match the true data generation but performance was not statistically different from the model that did match the true data generation; all such cases involved a multi-source model performing on-par with a single-source model when the true data were single-source.

\subsubsection*{Separate normal data}

Tables \ref{tab:normal_misclass_ld} and \ref{tab:normal_correlation_ld} give the misclassification rate and correlation with the true discriminating signal, respectively, for the low dimensional simulation with data generated from separate normal distributions.  In all cases, the model that best matched the data generation scenario performed the best for both measures.  The benefits of using the correct multiway structure tended to be more dramatic than that for matching the multi-source structure; the relative differences are particularly large for the correlations shown in Table~\ref{tab:normal_correlation_ld}. 


\subsubsection*{Continuous outcome, low dimension (N=100, P1=3, P2=3, d=5)}

Tables \ref{tab:continuous_rse_ld} and \ref{tab:continuous_correlation_ld} give the relative squared error and correlation with the true discriminating signal, respectively, for the low dimensional simulation for data generated with a continuous outcome.  In general, the model that best matched the data generation scenario performed the best for both measures, though some cases saw models that matched the true data structure failing to outperform models that did not. 
Interestingly, the rank 2 models closely match the performance of the rank 1 model (even if it is misspecified) but the full rank model performs much worse under low rank structure.  The benefits of using the correct multiway structure (Rank 2, 1, or full rank) tended to be more dramatic than that for matching the multi-source structure; the relative differences are particularly large for the correlations shown in Table~\ref{tab:normal_correlation_ld}. 


\subsubsection*{Binary outcome, high dimension (N=20, P1=100, P2=100, d=2)}

\subsubsection*{Probit generated data}

Tables \ref{tab:probit_misclass_hd} and \ref{tab:probit_correlation_hd} give the misclassification rate and correlation with the true discriminating signal, respectively, for the high dimensional simulation with probit generated data. In all cases, the model that best matched the data generation scenario performed the best for both measures, though the performance was not always statistically significant in outperforming other models. In particular, the misclassification rates observed for all models were close to 0.5, indicating performance only marginally better than random guessing. This demonstrates the challenge in fitting predictive models to HDLSS data when the distribution of the data does not depend on the outcome.    The correlation results were also not very strong, though they do more clearly indicate better performance from the models that match the true data generation.

\subsubsection*{Normal generated data}

Tables \ref{tab:normal_misclass_hd} and \ref{tab:normal_correlation_hd} give the misclassification rate and correlation with the true discriminating signal, respectively, for the  high-dimensional simulation with data generated from separate normal distributions.  In general the misclassification rates are much better here than they are in the high-dimensional probit scenario.  In all cases, the model that best matched the data generation scenario performed the best for both measures.  The benefits of using the correct multiway structure (Rank 1 or full rank) tended to be more dramatic than that for matching the multi-source structure; the relative differences are particularly large for the correlations shown in Table~\ref{tab:normal_correlation_hd}.

\subsubsection*{Continuous outcome, high dimension (N=20, P1=100, P2=100, d=2)}

Tables \ref{tab:continuous_rse_hd} and \ref{tab:continuous_correlation_hd} give the relative squared error and correlation with the true discriminating signal, respectively, for the high dimensional simulation with probit generated data. In all cases, the model that best matched the data generation scenario performed the best for both measures, though the performance was not statistically significant in outperforming other models. In particular, the relative squared errors observed for all models were rather close to 1, indicating performance that is only marginally beneficial. The correlation results were also not very strong, though they more clearly indicate better performance from the models that match the true data generation.

\subsection{Application to multi-omic iron deficiency}
\label{sec:Application}

We applied our method of multi-source, multi-way Bayesian probit regression to our motivating data on iron deficiency in an infant rhesus monkey model and assessed our ability to discriminate between ID and iron sufficient (IS) infants based on the serum proteomic and metabolomic profiles measured at two time points (4 and 6 months after birth). In this model, infants destined to develop ID show evidence of ID (changes in serum iron indices and lower reticulocyte hemoglobin content) at 4 months, with iron deficiency anemia (lower hemoglobin and mean corpuscular volume) seen at 6 months \citep{lubach2006preconception,coe2013optimal,rao2018metabolomic,sandri2020early,sandri2021correcting,sandri2022multiomic}. Proteomic and metabolomic changes in serum are seen in the preanemic and anemic periods \citep{sandri2022multiomic}. After routine pre-processing data were available for 227 metabolites and 205 proteins for $6$ ID and $6$ IS monkeys. We used the relatively non-informative prior to infer the variances of the proteomic and metabolomic coefficients, $\tau_1^2$ and $\tau_2^2$:  an Inverse Gamma distribution with parameters $\alpha = 1, \beta = 0.1$. 

\begin{figure}[h]
\centering
\includegraphics[width=\textwidth]{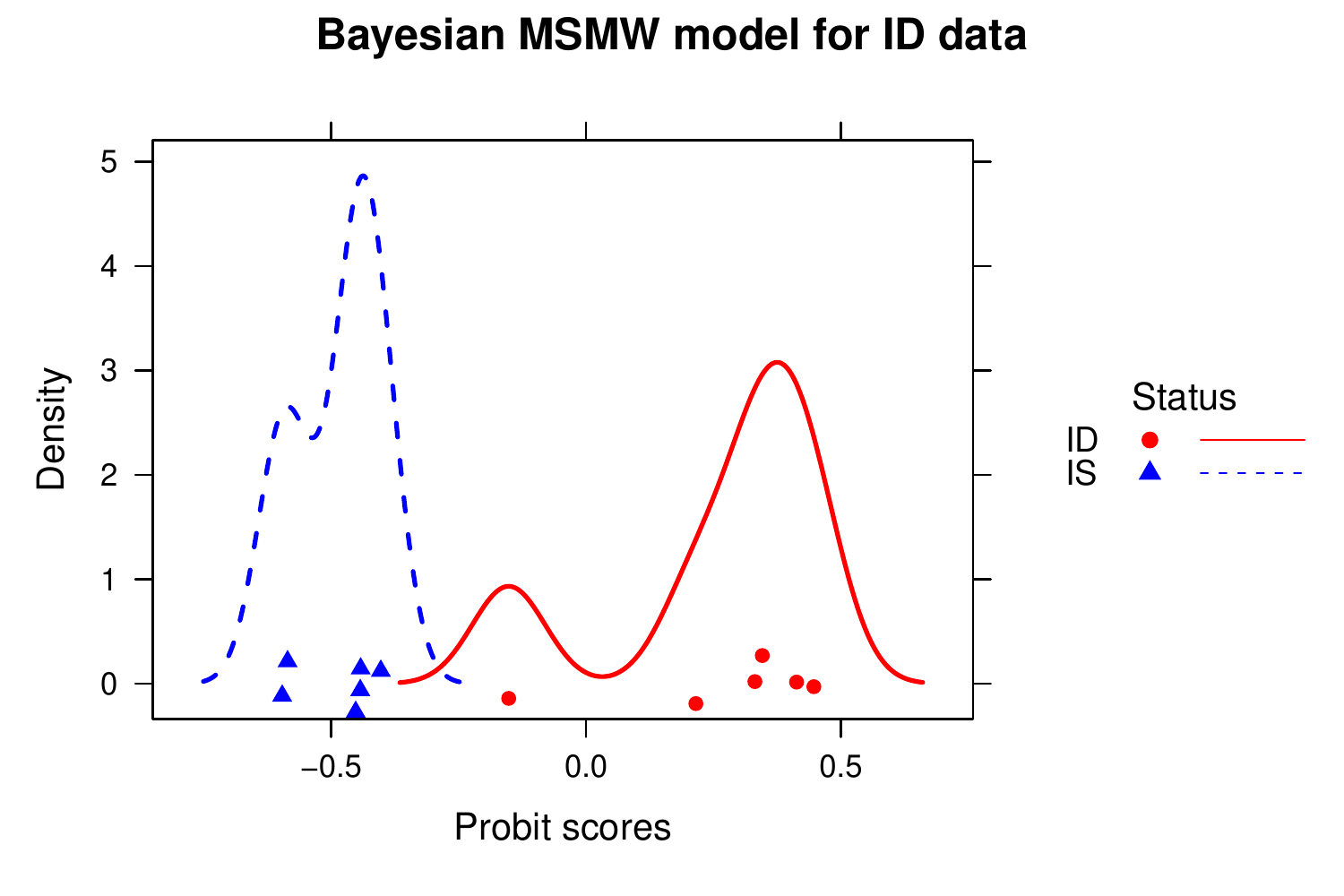}
\caption{Probit scores from applying MSMW model to motivating data under LOOCV.}
\label{fig:Scores}
\end{figure}

\begin{figure}[h]
\centering
\includegraphics[scale=1]{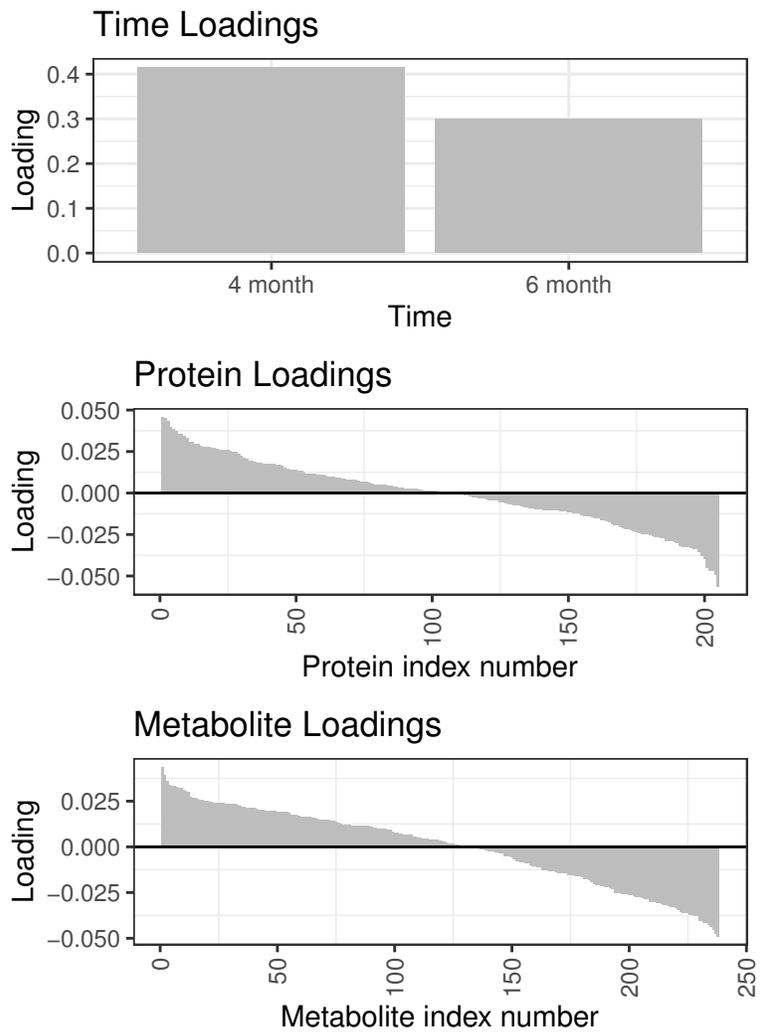}
\caption{Factor loadings from applying MSMW model to motivating data.}
\label{fig:loadings}
\end{figure}

We assessed the estimated probabilities of class membership under leave-one-out cross validation (LOOCV) using the rank-1 model, for which the posterior predictive probability for a held-out infant is inferred given the remaining $N-1$ infants. The plot of these probabilities demonstrated our model's ability to achieve perfect separation between the ID and IS samples in the estimated class probabilities (Figure \ref{fig:Scores}). We also examined the loadings for the individual proteins, individual metabolites, and each time point (Figure \ref{fig:loadings}).  The proteomic and metabolomic loadings both show several biomarkers that are positively and negatively associated with ID; moreover, the loadings have  similar scales between the two data sources, with $\tau_1=0.200$ (proteomics) and $\tau_2=0.238$ (metabolomics) indicating that the signal discriminating ID from IS infants is of similar size.    


To assess potential benefits of our approach, we compared the t-statistic for the difference in probit scores between the IS and ID groups under LOOCV to analogous approaches that do not account for multi-source or multi-way structure. Table~\ref{tab:realdatass} shows the resulting values for a multi-way (i.e., rank 1) or non multi-way (i.e., full rank) model using (1) only the metabolite data, (2) only the proteomic data, or (3) both data sources.  In all cases the multi-way approach performs better, suggesting that the metabolomic and proteeomic profiles discrimninating ID from IS infants are similar at 4-months and 6-months, and we can improve power by accounting for this structure.  Moreover, the chosen multi-way model with both sources outperforms others with a t-statistic of $7.235$, suggesting that the metabolites and proteins have complementary information and we can improve performance by combining them in a single model.  Moreover, an analogous approach that did not model the source variances separately acheived a small t-statistic of $5.521$, suggesting an advantage to accounting for heterogeneity between the sources.       



\section{Discussion}
\label{sec4}

We have proposed a Bayesian linear model that can predict a binary or a continuous outcome using data that are both multi-source and multi-way, with any number of sources or dimensions. Both the simulation and data analysis results have shown that the proposed MSMW model can improve classification accuracy and reduce MSE when the underlying data have MSMW structure. However, the performance of any given approach depends on the conditions that the data were generated, such as the true rank of the underlying signal or whether different sources have different signal variances. Thus, practical data applications of this model may require applying different versions of the method and comparing their performance.  In this article we have focused on three-way arrays ($N \times P \times D$), however, extensions to high-order arrays are straightforward, for which the coefficients array will take the form of a CP decomposition \citep{zhou2013tensor,guo2021multiway}.

\section{Software}
\label{sec5}

Software in the form of R code, together with a  sample
input data set and complete documentation is available at \url{https://github.com/BiostatsKim/BayesMSMW}.



\section*{Acknowledgments}
This work was supported by the National Institute of General Medical Sciences (NIGMS) grant R01-GM130622. Funding for the data application in Section~\ref{sec:Application} was also provided by grants from the National Institute of Health/Eunice Kennedy Shriver National Institute of Child Health and Development [HD089989, HD080201, HD057064 and HD39386].  



\bibliographystyle{biorefs}
\bibliography{refs2}

\newpage

\begin{table}[h]
\centering
\begin{tabular}{l c c c c c c}
\hline
\multicolumn{7}{c}{Misclassification: low-dimensional probit} \\
\hline
\multicolumn{1}{c}{} & \multicolumn{2}{c}{Rank: 2} & \multicolumn{2}{c}{Rank: 1} & \multicolumn{2}{c}{Full rank}\\
\cline{1-7}
\multicolumn{1}{c}{Model} & MS: Yes & MS: No & MS: Yes & MS: No & MS: Yes & MS: No \\
  \hline
Rank2,MS & \textbf{0.214} & \textbf{0.217} & 0.264 & 0.229 & 0.194 & 0.222 \\ 
  Rank2,SS & 0.216 & \textbf{0.216} & 0.264 & 0.227 & 0.198 & 0.221 \\ 
  Rank1,MS & 0.225 & 0.244 & \textbf{0.251} & \textbf{0.210} & 0.249 & 0.285 \\ 
  Rank1,SS & 0.224 & 0.243 & 0.252 & \textbf{0.210} & 0.250 & 0.284 \\ 
  FullRank,MS & 0.233 & 0.241 & 0.293 & 0.266 & \textbf{0.174} & 0.170 \\ 
  FullRank,SS & 0.248 & 0.241 & 0.300 & 0.265 & 0.196 & \textbf{0.168} \\ 
   \hline
\end{tabular}
\caption{Test misclassification rate for low-dimensional probit scenario.} 
\label{tab:probit_misclass_ld}
\end{table}

\begin{table}[h]
\centering
\begin{tabular}{l c c c c c c}
\hline
\multicolumn{7}{c}{Correlations: low-dimensional probit} \\
\hline
\multicolumn{1}{c}{} & \multicolumn{2}{c}{Rank: 2} & \multicolumn{2}{c}{Rank: 1} & \multicolumn{2}{c}{Full rank}\\
\cline{1-7}
\multicolumn{1}{c}{Model} & MS: Yes & MS: No & MS: Yes & MS: No & MS: Yes & MS: No \\
  \hline
  Rank2,MS & \textbf{0.750} & 0.739 & 0.620 & 0.710 & 0.131 & 0.135 \\ 
  Rank2,SS & 0.748 & \textbf{0.740} & 0.620 & 0.713 & 0.130 & 0.136 \\ 
  Rank1,MS & 0.730 & 0.690 & \textbf{0.642} & \textbf{0.737} & 0.117 & 0.094 \\ 
  Rank1,SS & 0.730 & 0.691 & 0.639 & \textbf{0.737} & 0.116 & 0.095 \\ 
  FullRank,MS & 0.096 & 0.109 & 0.085 & 0.106 & \textbf{0.837} & 0.857 \\ 
  FullRank,SS & 0.094 & 0.110 & 0.083 & 0.109 & 0.805 & \textbf{0.859} \\ 
   \hline
\end{tabular}
\caption{Correlation with true coefficients for the low-dimensional probit scenario.} 
\label{tab:probit_correlation_ld}
\end{table}

\begin{table}[h]
\centering
\begin{tabular}{l c c c c c c}
\hline
\multicolumn{7}{c}{Misclassification: low-dimensional separate normal} \\
\hline
\multicolumn{1}{c}{} & \multicolumn{2}{c}{Rank: 2} & \multicolumn{2}{c}{Rank: 1} & \multicolumn{2}{c}{Full rank}\\
\cline{1-7}
\multicolumn{1}{c}{Model} & MS: Yes & MS: No & MS: Yes & MS: No & MS: Yes & MS: No \\
\hline
  Rank2,MS & \textbf{0.111} & \textbf{0.034} & 0.246 & 0.134 & 0.176 & 0.118 \\ 
  Rank2,SS & \textbf{0.111} & \textbf{0.033} & 0.245 & 0.132 & 0.175 & 0.115 \\ 
  Rank1,MS & 0.117 & 0.043 & \textbf{0.240} & 0.129 & 0.200 & 0.155 \\ 
  Rank1,SS & 0.116 & 0.042 & \textbf{0.241} & \textbf{0.128} & 0.200 & 0.154 \\ 
  FullRank,MS & 0.115 & 0.036 & 0.258 & 0.146 & \textbf{0.168} & 0.092 \\ 
  FullRank,SS & 0.117 & 0.035 & 0.261 & 0.145 & 0.174 & \textbf{0.091} \\ 
   \hline
\end{tabular}
\caption{Test misclassification rate for the low-dimensional separate normal scenario.} 
\label{tab:normal_misclass_ld}
\end{table}

\begin{table}[h]
\centering
\begin{tabular}{l c c c c c c}
\hline
\multicolumn{7}{c}{Correlations: low-dimensional separate normal} \\
\hline
\multicolumn{1}{c}{} & \multicolumn{2}{c}{Rank: 2} & \multicolumn{2}{c}{Rank: 1} & \multicolumn{2}{c}{Full rank}\\
\cline{1-7}
\multicolumn{1}{c}{Model} & MS: Yes & MS: No & MS: Yes & MS: No & MS: Yes & MS: No \\
  \hline
  Rank2,MS & \textbf{0.876} & 0.887 & 0.764 & 0.877 & 0.164 & 0.091 \\ 
  Rank2,SS & \textbf{0.873} & \textbf{0.897} & 0.770 & 0.885 & 0.170 & 0.098 \\ 
  Rank1,MS & 0.846 & 0.844 & \textbf{0.796} & 0.901 & 0.167 & 0.087 \\
  Rank1,SS & 0.848 & 0.845 & \textbf{0.796} & \textbf{0.905} & 0.166 & 0.090 \\
  FullRank,MS & 0.196 & 0.155 & 0.111 & 0.125 & \textbf{0.832} & 0.845 \\ 
  FullRank,SS & 0.192 & 0.160 & 0.106 & 0.125 & 0.814 & \textbf{0.853} \\ 
   \hline
\end{tabular}
\caption{Correlation with true coefficients for the low-dimensional separate normal scenario.} 
\label{tab:normal_correlation_ld}
\end{table}

\begin{table}[h]
\centering
\begin{tabular}{l c c c c c c}
\hline
\multicolumn{7}{c}{Relative Squared Error: low-dimensional continuous} \\
\hline
\multicolumn{1}{c}{} & \multicolumn{2}{c}{Rank: 2} & \multicolumn{2}{c}{Rank: 1} & \multicolumn{2}{c}{Full rank}\\
\cline{1-7}
\multicolumn{1}{c}{Model} & MS: Yes & MS: No & MS: Yes & MS: No & MS: Yes & MS: No \\
\hline
  Rank2,MS & 0.742 & 0.446 & 0.719 & 0.616 & \textbf{0.500} & 0.468 \\ 
  Rank2,SS & \textbf{0.740} & \textbf{0.444} & 0.716 & 0.613 & 0.501 & 0.464 \\ 
  Rank1,MS & 0.758 & 0.550 & \textbf{0.700} & \textbf{0.602} & 0.654 & 0.664 \\ 
  Rank1,SS & 0.757 & 0.549 & \textbf{0.700} & \textbf{0.601} & 0.654 & 0.661 \\ 
  FullRank,MS & 0.811 & 0.487 & 0.801 & 0.689 & \textbf{0.485} & 0.287 \\ 
  FullRank,SS & 0.784 & 0.473 & 0.777 & 0.668 & \textbf{0.485} & \textbf{0.286} \\ 
   \hline
\end{tabular}
\caption{Mean relative squared prediction error on test data for the low-dimensional continuous scenario.} 
\label{tab:continuous_rse_ld}
\end{table}

\begin{table}[h]
\centering
\begin{tabular}{l c c c c c c}
\hline
\multicolumn{7}{c}{Correlations: low-dimensional continuous} \\
\hline
\multicolumn{1}{c}{} & \multicolumn{2}{c}{Rank: 2} & \multicolumn{2}{c}{Rank: 1} & \multicolumn{2}{c}{Full rank}\\
\cline{1-7}
\multicolumn{1}{c}{Model} & MS: Yes & MS: No & MS: Yes & MS: No & MS: Yes & MS: No \\
\hline
  Rank2,MS & \textbf{0.703} & \textbf{0.873} & 0.751 & \textbf{0.810} & 0.222 & 0.099 \\ 
  Rank2,SS & 0.702 & \textbf{0.874} & 0.751 & \textbf{0.811} & 0.222 & 0.101 \\ 
  Rank1,MS & 0.667 & 0.792 & \textbf{0.774} & 0.819 & 0.191 & 0.092 \\ 
  Rank1,SS & 0.666 & 0.792 & \textbf{0.772} & \textbf{0.820} & 0.192 & 0.079 \\ 
  FullRank,MS & 0.137 & 0.153 & 0.062 & 0.141 & \textbf{0.882} & \textbf{0.943} \\ 
  FullRank,SS & 0.136 & 0.152 & 0.060 & 0.142 & 0.874 & \textbf{0.944} \\ 
   \hline
\end{tabular}
\caption{Correlation with true coefficients for the low-dimensional continuous scenario.} 
\label{tab:continuous_correlation_ld}
\end{table}

\begin{table}[h]
\centering
\begin{tabular}{l c c c c }
\hline
\multicolumn{5}{c}{Misclassification: high-dimensional probit} \\
\hline
\multicolumn{1}{c}{} & \multicolumn{2}{c}{Rank: 1} & \multicolumn{2}{c}{Full rank}\\
\cline{1-5}
\multicolumn{1}{c}{Model} & MS: Yes & MS: No & MS: Yes & MS: No \\
  \hline
  Rank1,MS & \textbf{0.444} & \textbf{0.453} & 0.451 & 0.448 \\ 
  Rank1,SS & \textbf{0.448} & \textbf{0.454} & 0.454 & 0.447 \\ 
  FullRank,MS & \textbf{0.446} & \textbf{0.454} & \textbf{0.448} & \textbf{0.444} \\ 
  FullRank,SS & 0.449 & 0.452 & 0.453 & \textbf{0.443} \\ 
   \hline
\end{tabular}
\caption{Test misclassification rates for the high-dimensional probit scenario.}
\label{tab:probit_misclass_hd}
\end{table}

\begin{table}[h]
\centering
\begin{tabular}{l c c c c}
\hline
\multicolumn{5}{c}{Correlations: high-dimensional probit} \\
\hline
\multicolumn{1}{c}{} & \multicolumn{2}{c}{Rank: 1} & \multicolumn{2}{c}{Full rank}\\
\cline{1-5}
\multicolumn{1}{c}{Model} & MS: Yes & MS: No & MS: Yes & MS: No \\
  \hline
  Rank1,MS & \textbf{0.142} & 0.134 & 0.081 & 0.068 \\ 
  Rank1,SS & \textbf{0.147} & \textbf{0.148} & 0.079 & 0.070 \\ 
  FullRank,MS & 0.084 & 0.080 & \textbf{0.181} & 0.163 \\ 
  FullRank,SS & 0.080 & 0.082 & 0.172 & \textbf{0.172} \\ 
   \hline
\end{tabular}
\caption{Correlation with true coefficients for the high-dimensional probit scenario.} 
\label{tab:probit_correlation_hd}
\end{table}

\begin{table}[h]
\centering
\begin{tabular}{l c c c c}
\hline
\multicolumn{5}{c}{Misclassification: high-dimensional separate normal} \\
\hline
\multicolumn{1}{c}{} & \multicolumn{2}{c}{Rank: 1} & \multicolumn{2}{c}{Full rank}\\
\cline{1-5}
\multicolumn{1}{c}{Model} & MS: Yes & MS: No & MS: Yes & MS: No \\
  \hline
Rank1,MS & \textbf{0.203} & \textbf{0.178} & 0.172 & 0.068 \\ 
  Rank1,SS & 0.213 & \textbf{0.178} & 0.199 & 0.061 \\ 
  FullRank,MS & 0.215 & 0.198 & \textbf{0.157} & 0.054 \\ 
  FullRank,SS & 0.228 & 0.196 & 0.188 & \textbf{0.052} \\ 
   \hline
\end{tabular}
\caption{Test misclassification rates for the high-dimensional separate normal scenario.} 
\label{tab:normal_misclass_hd}
\end{table}

\begin{table}[h]
\centering
\begin{tabular}{l c c c c}
\hline
\multicolumn{5}{c}{Correlation: high-dimensional separate normal data} \\
\hline
\multicolumn{1}{c}{} & \multicolumn{2}{c}{Rank: 1} & \multicolumn{2}{c}{Full rank}\\
\cline{1-5}
\multicolumn{1}{c}{Model} & MS: Yes & MS: No & MS: Yes & MS: No \\
  \hline
Rank1,MS & \textbf{0.468} & 0.479 & 0.202 & 0.251 \\ 
  Rank1,SS & 0.423 & \textbf{0.492} & 0.187 & 0.253 \\ 
  FullRank,MS & 0.226 & 0.200 & \textbf{0.487} & 0.534 \\ 
  FullRank,SS & 0.200 & 0.205 & 0.421 & \textbf{0.546} \\ 
   \hline
\end{tabular}
\caption{Correlation with true coefficients for the high-dimensional separate normal scenario.} 
\label{tab:normal_correlation_hd}
\end{table}

\begin{table}[h!]
\centering
\begin{tabular}{l c c c c}
\hline
\multicolumn{5}{c}{Relative squared error: high-dimensional continuous} \\
\hline
\multicolumn{1}{c}{} & \multicolumn{2}{c}{Rank: 1} & \multicolumn{2}{c}{Full rank}\\
\cline{1-5}
\multicolumn{1}{c}{Model} & MS: Yes & MS: No & MS: Yes & MS: No \\
  \hline
  Rank1,MS & \textbf{0.972} & \textbf{0.965} & \textbf{0.957} & 0.984 \\ 
  Rank1,SS & 0.977 & \textbf{0.969} & 0.964 & 0.975 \\ 
  FullRank,MS & 0.984 & 0.972 & \textbf{0.952} & 0.963 \\ 
  FullRank,SS & 0.983 & \textbf{0.968} & \textbf{0.955} & \textbf{0.958} \\ 
   \hline
\end{tabular}
\caption{Mean relative squared prediction error on test data for the high-dimensional continuous scenario.} 
\label{tab:continuous_rse_hd}
\end{table}

\begin{table}[h!]
\centering
\begin{tabular}{l c c c c}
\hline
\multicolumn{5}{c}{Correlation Results: Continuous data} \\
\hline
\multicolumn{1}{c}{} & \multicolumn{2}{c}{Rank: 1} & \multicolumn{2}{c}{Full rank}\\
\cline{1-5}
\multicolumn{1}{c}{Model} & MS: Yes & MS: No & MS: Yes & MS: No \\
  \hline
  Rank1,MS & \textbf{0.179} & \textbf{0.185} & 0.096 & 0.091 \\ 
  Rank1,SS & 0.169 & \textbf{0.180} & 0.090 & 0.099 \\ 
  FullRank,MS & 0.072 & 0.100 & \textbf{0.223} & 0.211 \\ 
  FullRank,SS & 0.070 & 0.102 & \textbf{0.215} & \textbf{0.218} \\ 
   \hline
\end{tabular}
\caption{Correlation with true coefficients for the high-dimensional continuous scenario.} 
\label{tab:continuous_correlation_hd}
\end{table}



\begin{table}[h]
\centering
\begin{tabular}{l c c}
\hline
\multicolumn{3}{c}{t-test statistics} \\
\hline
\multicolumn{1}{c}{} & \multicolumn{1}{c}{Multi-way} & \multicolumn{1}{c}{Non-multi-way}\\
  \hline
  Metabolites only & \textbf{5.773} & 4.407   \\ 
  Proteins only  & \textbf{5.182} & 3.935  \\ 
  All data  & \textbf{7.235} & 4.741  \\ 

   \hline
\end{tabular}
\caption{Test statistics from using two-sample t tests to evaluate separation between ID and IS samples achieved by different models under LOOCV.}
\label{tab:realdatass}
\end{table}

\end{document}